\documentclass{phyeauth}
\usepackage{graphicx}
\usepackage{bm}

\begin{document}

\begin{frontmatter}
  \title{Angular magnetoresistance oscillations in bilayers in tilted
    magnetic fields}
  
  \author{Victor M. Yakovenko\thanksref{VMY}}, 
  \author{Benjamin K. Cooper} 
  
  \address{Condensed Matter Theory Center, Department of Physics,
    University of Maryland, College Park, MD 20742--4111, USA}
  
  \thanks[VMY]{ Corresponding author.  E-mail: yakovenk@umd.edu}


\begin{abstract}
  Angular magnetoresistance oscillations (AMRO) were originally
  discovered in organic conductors and then found in many other
  layered metals.  It should be possible to observe AMRO to
  semiconducting bilayers as well.  Here we present an intuitive
  geometrical interpretation of AMRO as the Aharonov-Bohm interference
  effect, both in real and momentum spaces, for balanced and
  imbalanced bilayers.  Applications to the experiments with bilayers
  in tilted magnetic fields in the metallic state are discussed.  We
  speculate that AMRO may be also observed when each layer of the
  bilayer is in the composite-fermion state.
\end{abstract}

\end{frontmatter}

The so-called angular magnetoresistance oscillations (AMRO) were
originally discovered in the quasi-two-dimensional (Q2D) organic
conductors of the (BEDT-TTF)$_2$X family \cite{Kartsovnik88,Kajita}.
Upon rotation of a magnetic field $\bm B$, electrical resistivity
oscillates periodically in $\tan\theta$, where $\theta$ is the angle
between $\bm B$ and the normal to the layers.  The oscillations are
very strong and the most pronounced in the interlayer resistivity
$\rho_z$.  AMRO are distinct from the Shubnikov-de Haas (SdH)
oscillations, where resistivity oscillates as a function of the
magnetic field magnitude for a fixed orientation.  In AMRO,
resistivity has maxima at certain angles $\theta$, often called the
``magic angles'', that are independent of the magnetic field
magnitude.  AMRO typically persist to substantially higher
temperatures than the SdH oscillations, so the two effects can be
clearly separated experimentally.  Theory explained that the period of
AMRO in $\tan\theta$ is inversely proportional to $k_Fd$, where $d$ is
the interlayer distance, and $k_F$ is the in-plane Fermi wave vector.
Thus, AMRO can be utilized to determine $k_F$ and to map out Fermi
surfaces of Q2D materials with anisotropic $k_F$.  This was done first
in $\beta$-(BEDT-TTF)$_2$IBr$_2$ \cite{Kartsovnik91}, and then in a
variety of organic conductors (see reviews
\cite{Kartsovnik96,Wosnitza96,Singleton00}).  AMRO were also observed
in many other layered materials, such as intercalated graphite
\cite{Iye}, $\rm Sr_2RuO_4$ \cite{Ohmichi}, $\rm Tl_2Ba_2CuO_6$
\cite{Dragulescu,Hussey}, and the GaAs superlattices
\cite{GaAS91,GaAs98,GaAs01}.

The first theory of AMRO was presented by Yamaji \cite{Yamaji}, who
pointed out that the amplitude of the SdH oscillations should be
maximal at the magic angles determined by zeroes of the Bessel
function $J_0(k_Fd\tan\theta)$.  Yagi \emph{et al.}\ \cite{Yagi}
calculated angular oscillations of the interlayer conductivity
$\sigma_z(\theta)$ from the Boltzmann equation using semiclassical
electron trajectories on the cylindrical 3D Fermi surface.  It was
assumed that a periodic crystal with many layers and a 3D Fermi
surface is necessary for observation of AMRO.  However, it was also
recognized \cite{Yagi} that AMRO exist already in the limit of
infinitesimal interlayer tunneling amplitude $t_\perp\to0$.  Using the
Landau wave functions, Kurihara \cite{Kurihara92,Kurihara93} and
Yoshioka \cite{Yoshioka} calculated the effective interlayer tunneling
amplitude $\tilde t_\perp$ in a tilted magnetic field and found
angular oscillations in $\tilde t_\perp(\theta)$.  Then McKenzie and
Moses \cite{McKenzie,Moses} explicitly demonstrated that electron
tunneling between just two layers shows AMRO due to interference of
the gauge phase differences between the layers.  These ideas were
further developed by Osada \emph{et al.}\ for Q2D and Q1D materials
\cite{Osada02,Osada03}.

Meanwhile, semiconducting bilayers were studied experimentally in
parallel \cite{Eisenstein,Simmons93} and tilted
\cite{Boebinger91,Harff} magnetic fields.  On the theory side, Hu and
MacDonald \cite{MacDonald92} calculated $\tilde t_\perp$ in a tilted
field using the Landau wave functions, and Lyo \emph{et al.}\ 
\cite{Lyo93,Lyo98,Simmons98} studied conductivity using the Kubo
formula.  They found vanishing $\tilde t_\perp$ for certain angles
$\theta$ \cite{MacDonald92} and oscillatory dependence of $\sigma_z$
on the magnetic field component $B_\|$ parallel to the layers for a
fixed perpendicular component $B_\perp$ \cite{Lyo98}.  However, these
papers (also \cite{Fertig02}) focused on the low Landau filling
factors, whereas Q2D metals were studied for the high filling factors,
so a relation between AMRO in these two classes of materials was not
recognized.

\begin{figure}
\begin{centering}
  \includegraphics[width=0.46\linewidth]{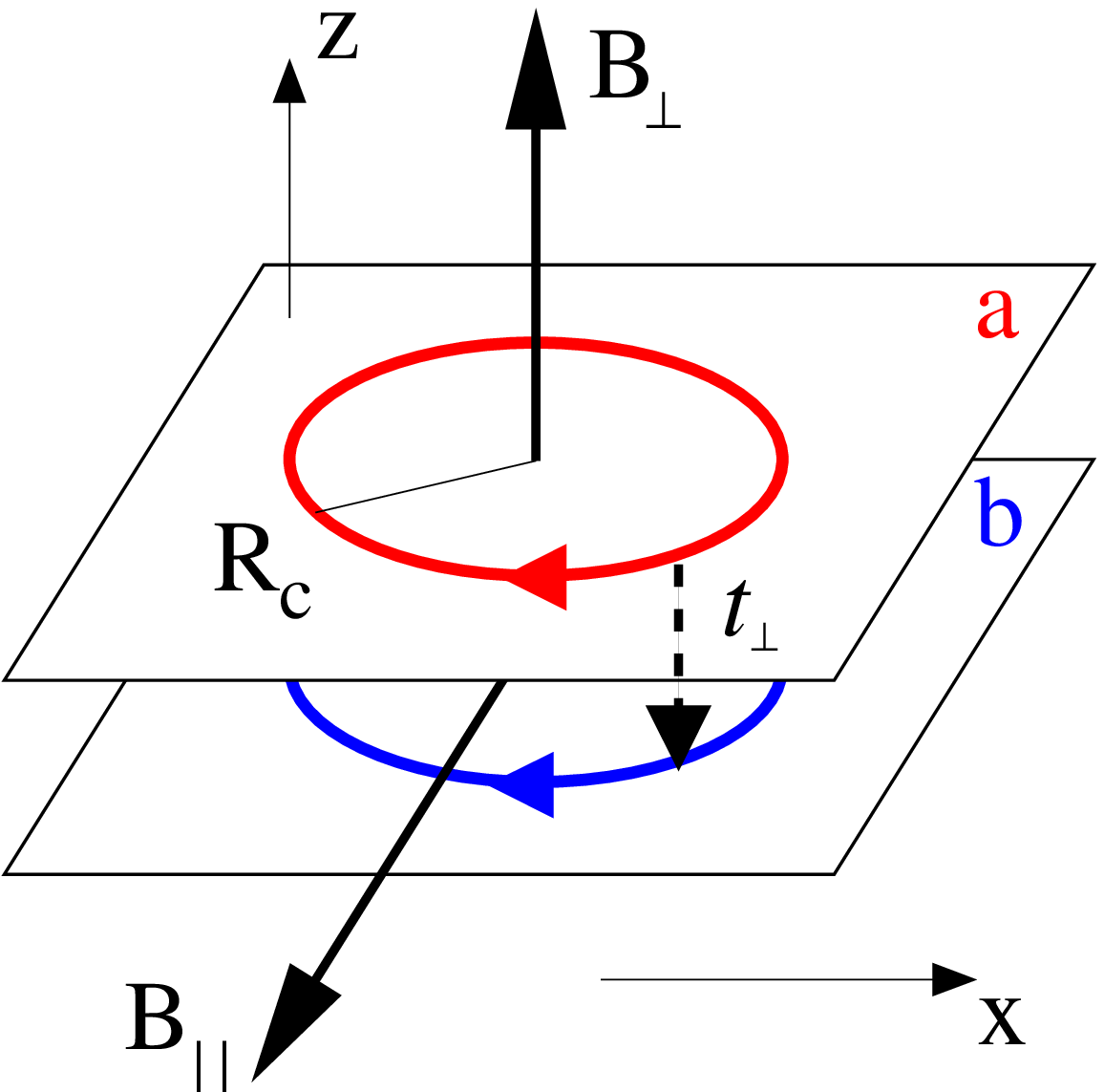} \hfill
  \includegraphics[width=0.46\linewidth]{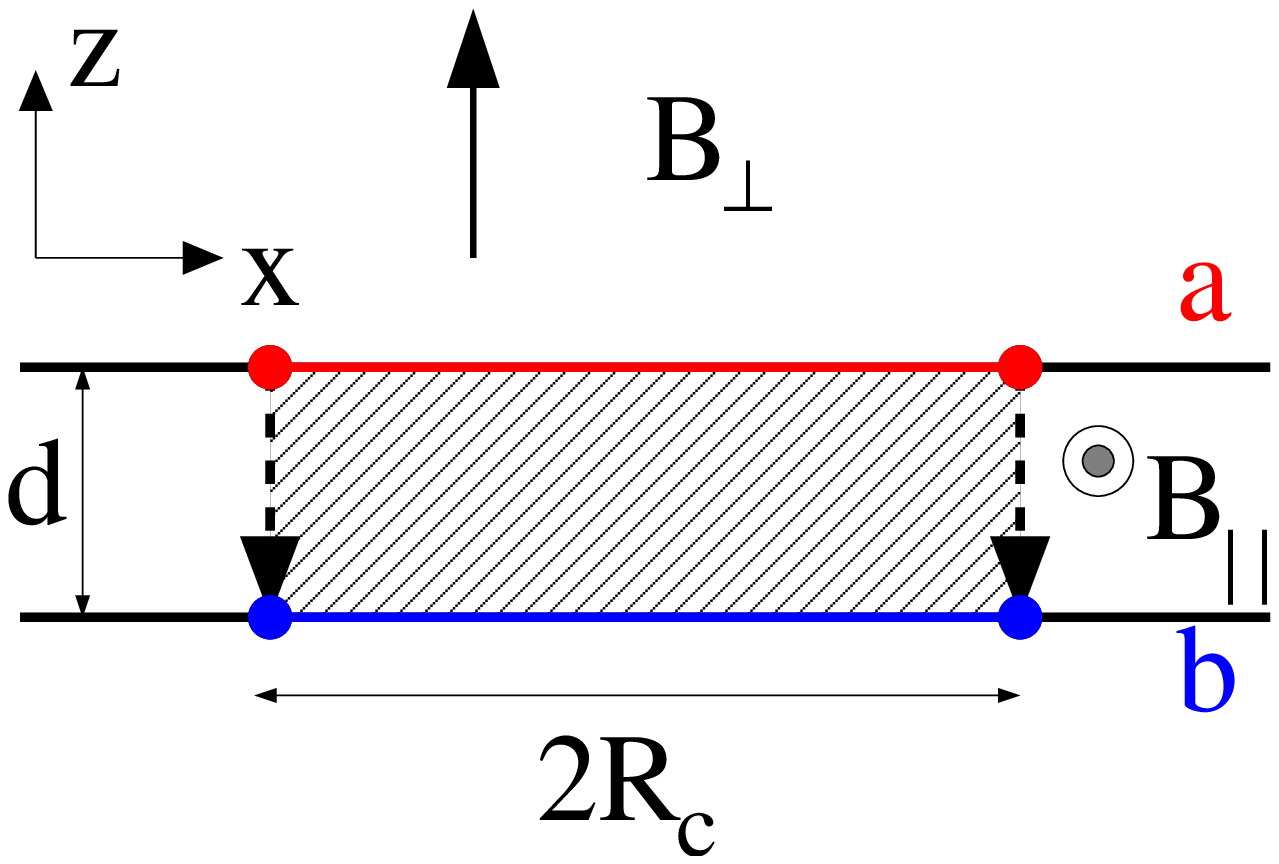}
\caption{(a) Tunneling geometry for a bilayer in a tilted magnetic 
  field $\bm B$.  The circles represent the quasiclassical cyclotron
  orbits.  (b) View of the bilayer along the layers.  Magnetic flux
  through the shaded area produces oscillations of the effective
  interlayer tunneling amplitude.}
\label{fig:tunneling}
\end{centering}
\end{figure}

In this paper, we would like to make a connection between AMRO in
layered metals and semiconducting bilayers.  We present an intuitive
geometrical interpretation of AMRO as the Aharonov-Bohm effect, both
in real and momentum spaces.  We start with the density-balanced
bilayers, where both layers have the same Fermi surfaces, and then
generalize to the density-imbalanced bilayers with different Fermi
surfaces.  We also speculate that it may be possible to observe AMRO
when the layers are in the composite-fermion state and to use AMRO for
investigation of such a state.  We hope that fresh insight from the
organic conductors community will be stimulating for further studies
of oscillatory phenomena in semiconductor bilayers (for the Q1D case
see \cite{Yakovenko}).

The bilayer geometry is shown in Fig.\ \ref{fig:tunneling}.  Electron
tunneling between the layers a and b is described by the Hamiltonian
\begin{equation}
  \hat{H}_{\perp} = t_{\perp} \int \hat\psi_a^\dag({\bm r})\,
  \hat\psi_b({\bm r})\, e^{\frac{ieA_z({\bm r})d}{\hbar c}} d^2r\,
  + \rm{H.c.},
\label{eq:H_perp}
\end{equation}
where we have chosen the gauge $A_z=B_\|x$.  We will assume that the
interlayer tunneling amplitude $t_{\perp}$ is small compared with the
intralayer energy scales, so it can be treated as a perturbation.  We
will use a quasiclassical approximation to describe the in-plane
electron motion, assuming that the Landau filling factors are high
enough.  In the presence of $B_{\perp}$, electrons execute
quasiclassical cyclotron motion within the layers with the frequency
$\omega_c=eB_{\perp}/m$ and the radius $R_c=c\hbar k_F/eB_{\perp}$.
Here we used the Fermi wave vector $k_F$ in the formula for $R_c$,
because only the electrons at the Fermi surface are relevant for
conduction.  For balanced bilayers, $k_F$ is the same in both layers.

\begin{figure}
\begin{centering}
  \includegraphics[width=0.8\linewidth]{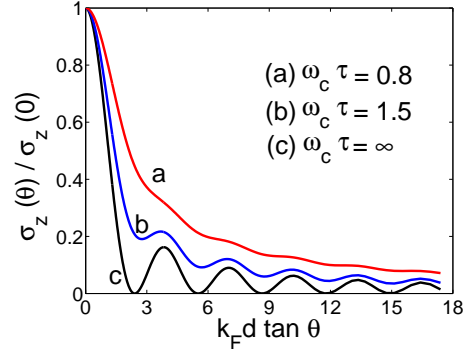}
\caption{Interlayer conductivity $\sigma_z$ calculated from Eq.\ 
  (\ref{eq:sigma-tau}) and plotted vs.\ $\tan\theta=B_\|/B_\perp$.}
\label{fig:bessel}
\end{centering}
\end{figure}

The gauge phase in Eq.\ (\ref{eq:H_perp}) leads to interference between
electron tunneling at different points along the cyclotron orbit, and
the effective tunneling amplitude $\tilde t_\perp$ is obtained by
phase averaging \cite{McKenzie,Moses}:
\begin{equation}
  \tilde t_\perp = t_\perp \left\langle 
  e^{\frac{ieB_\|x(t)d}{\hbar c}} \right\rangle_t
  = t_\perp J_0(k_Fd\tan\theta).
\label{eq:ttilde}
\end{equation}
Here the brackets represent averaging over time $t$ for the cyclotron
motion $x(t)=R_c\cos(\omega_c t)$, $J_0$ is the Bessel function, and
$\tan\theta=B_\|/B_\perp$.  Since the interlayer tunneling
conductivity $\sigma_z$ is proportional to $\tilde t_\perp^2$, Eq.\ 
(\ref{eq:ttilde}) gives
$\sigma_z(\theta)/\sigma_z(0)=J_0^2(k_Fd\tan\theta)$, which is shown
by the curve (c) in Fig.\ \ref{fig:bessel}.  From the asymptotic
expression $J_0(\xi)\propto\cos(\xi-\pi/4)/\sqrt{\xi}$, we find that
$\tilde t_\perp$ and $\sigma_z$ oscillate periodically in $\tan\theta$
and vanish at the ``magic angles''
\begin{equation}
  \frac{B_\|}{B_\perp}=\tan\theta_n=\frac{\pi(n-1/4)}{k_Fd},
\label{eq:theta_n}
\end{equation}
where $n$ is an integer.  This is the AMRO effect discussed in the
introduction.  In \cite{MacDonald92,Lyo98,Fertig02}, the effective
tunneling amplitude $\tilde t_\perp$ was obtained as a matrix element
of the Hamiltonian (\ref{eq:H_perp}) between the Landau wave functions
and expressed in terms of the Laguerre polynomials.  However, as
pointed out in Refs.\ \cite{Kurihara92,Kurihara93,Yoshioka}, the
Laguerre polynomials reduce to the Bessel function for the high Landau
levels, so the quasiclassical expression (\ref{eq:ttilde}) agrees with
the quantum calculation \cite{MacDonald92,Lyo98,Fertig02}.

Vanishing of $\tilde t_\perp$ at the magic angles not only results in
minima of $\sigma_z$, but also in disappearance of beating in the SdH
oscillations.  Generally, the symmetric and antisymmetric electron
states in a density-balanced bilayer are split in energy by $\tilde
t_\perp$, which results in two slightly different SdH frequencies.
However, at the magic angles, the energy split and the beating of the
SdH oscillations should disappear, because $\tilde t_\perp\to0$.  This
effect is observed in organic conductors \cite{Wosnitza96} and was
explained theoretically by Yamaji \cite{Yamaji}.  In bilayers, it was
observed \cite{Boebinger91} that the SdH beating period increases with
the increase of $B_\|$, in qualitative agreement with the argument
presented above.  However, the ratio $B_\|/B_\perp$ was not big enough
to reach a magic angle and to observe disappearance of the SdH
beating.

AMRO can be interpreted geometrically as a particular manifestation of
the Aharonov-Bohm effect.  Let us look at the bilayer along the
layers, as shown in Fig. \ref{fig:tunneling}b.  The gauge phase in
Eq.\ (\ref{eq:H_perp}) is proportional to the area contained between
the layers up to the point of electron tunneling.  The lines of the
length $2R_c$ represent the side view of the cyclotron orbits.
Electrons spend more time at the extremal turning points denoted as
the dots, which naturally define the shaded area $2R_cd$.  The
magnetic flux $\Phi$ through this area results in destructive
interference between electron tunneling at the opposite turning points
and vanishing of $\tilde t_\perp$ when $\Phi=2R_cdB_\|=\phi_0(n + C)$,
where $\phi_0=2\pi\hbar c/e$ is the flux quantum, and $C$ is an
appropriate constant.  Inserting the expression for $R_c$, we recover
Eq.\ (\ref{eq:theta_n}).  Notice that one dimension $d$ of the
Aharonov-Bohm area is fixed by the bilayer structure, but the other
dimension $2R_c$ is adjustable and is proportional to $B_\perp^{-1}$.
This results in the condition (\ref{eq:theta_n}) on the ratio of
$B_\|$ and $B_\perp$.

\begin{figure}
\begin{centering}
  \includegraphics[width=0.5\linewidth]{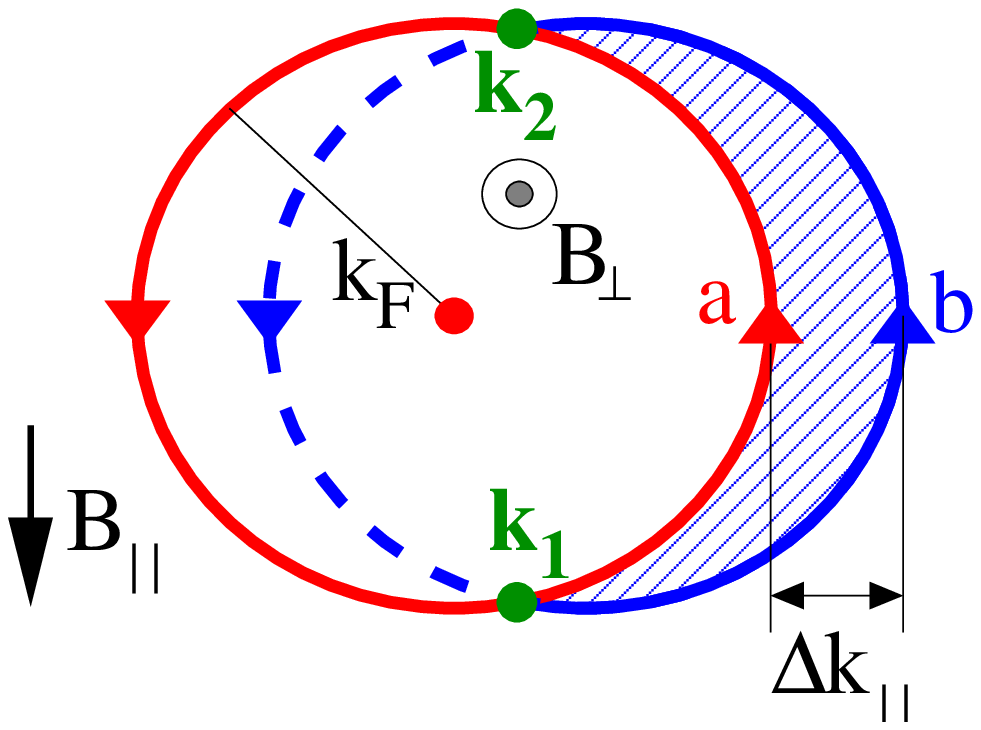} \hfill
  \includegraphics[width=0.4\linewidth]{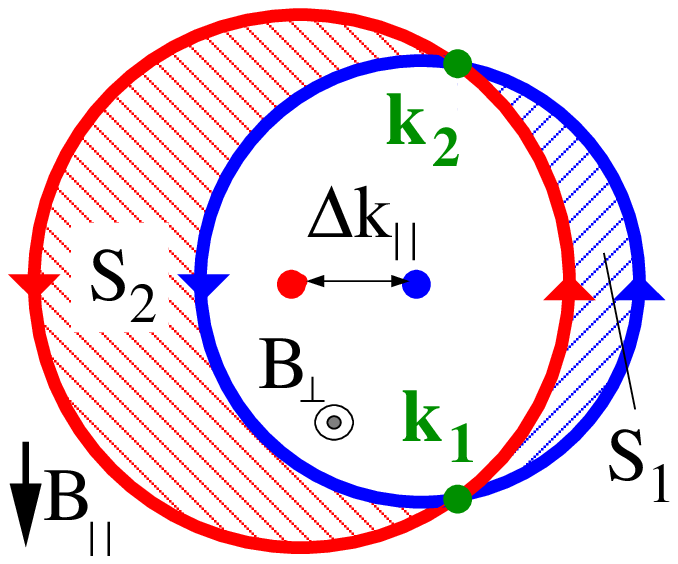}
\caption{Fermi surfaces of the layers in a  bilayer that is
  (a) balanced and (b) imbalanced.  The Fermi surfaces are displaced
  by $\Delta k_\|=eB_\|d/c$.  The magnetic flux associated with the
  shaded areas in momentum space causes oscillations of the effective
  interlayer tunneling amplitude.}
\label{fig:ABkspace}
\end{centering}
\end{figure}

AMRO can be also interpreted as a result of interference in the
momentum space, as illustrated in Fig.\ \ref{fig:ABkspace}.  Suppose
that only the $B_\|$ component is applied.  Then, according to Eq.\ 
(\ref{eq:H_perp}), the in-plane electron momentum changes by $\Delta
k_\|=eB_\|d/c\hbar$ upon tunneling between the layers
\cite{Eisenstein,Simmons93,Boebinger91}, so the Fermi surfaces of the
two layers are shifted relative to each other as shown in Fig.\ 
\ref{fig:ABkspace}.  Thus, electrons can tunnel only at the points
${\bm k}_1$ and ${\bm k}_2$, where the conservation laws of both
energy and momentum are satisfied.  When the $B_\perp$ component is
turned on, it causes interference between the two trajectories a and b
connecting the points ${\bm k}_1$ and ${\bm k}_2$.  The phase
difference between the two trajectories is proportional to the shaded
area $S$ between them in momentum space.  In the balanced case shown
in Fig.\ \ref{fig:ABkspace}a, $S\approx2k_F\Delta k_\|$, where we
assumed that $\Delta k_\|\ll k_F$, which is a typical condition for
Q2D metals.  The interference between the two momentum-space
trajectories is destructive when the condition
$B_\perp=\phi_0S/(n+C)(2\pi)^2$ is satisfied, which reproduces Eq.\ 
(\ref{eq:theta_n}).

In the imbalanced case, the interference oscillations develop between
the parallel trajectories that involve the momentum-space areas $S_1$
and $S_2$ in Fig.\ \ref{fig:ABkspace}b.  The frequencies of these
oscillations are given by the SdH-like formula
$B_\perp=\phi_0S_{1,2}/(n+C)(2\pi)^2$, where the areas $S_1$ and $S_2$
depend on $B_\|$.  Notice that these interference oscillations are
different from the SdH oscillations.  The later are the consequence of
the energy quantization originating from closed orbits, whereas the
former result from quantum interference between parallel orbits that
do not form a closed loop and do not produce energy quantization.
Magnetoresistance oscillations due to the momentum-space interference
are known in some metals \cite{Stark} and organic conductors
\cite{ClO4}.  The in-plane resistivity $\sigma_x$ of an imbalanced
bilayer in tilted magnetic fields was measured in Ref.\ \cite{Harff}.
The oscillations originating from the areas $S_1$ and $S_2$ can
probably be found in the Fourier spectrum shown in Fig.\ 4 of Ref.\ 
\cite{Harff}.  However, this paper focused only on the SdH
oscillations originating from closed orbits, but not on the
interference oscillations from parallel orbits.  In Fig.\ 3 of this
paper, one can recognize a pattern of lines at certain angles
$\tan\theta=B_\|/B_\perp$, which can be interpreted as observation of
AMRO.  It would be very interesting to measure the interlayer
conductivity $\sigma_z$, where the AMRO effect should be stronger than
in $\sigma_x$.

A finite lifetime $\tau$ of quasiparticles results in loss of phase
coherence, which can be described phenomenologically by an
exponentially decaying factor in the Kubo formula for $\sigma_z$
\cite{Yagi,McKenzie,Moses}:
\begin{equation}
  \sigma_z \propto t_{\perp}^2 \left\langle \int_t^{\infty}
  e^{\frac{i e B_{\|} d}{\hbar c} [x(t) - x(t^{\prime})]} 
  e^{- \frac{t^{\prime} - t}
  {\tau}} dt^{\prime} \right\rangle_t.
\label{eq:finitetau}
\end{equation}
Doing the integral in Eq.\ (\ref{eq:finitetau}), one finds
\cite{Yagi,McKenzie,Moses}
\begin{equation}
  \frac{\sigma_z({\bm B})}{\sigma_z(0)}=
  J_0^2(k_Fd\tan\theta) + 2\sum\limits_{j=1}^{\infty}
  \frac{J_j^2(k_Fd\tan\theta)}{1+(j\omega_c\tau)^2}.
\label{eq:sigma-tau}
\end{equation}
For $\omega_c\tau\gg1$, Eq.\ (\ref{eq:finitetau}) gives
$\sigma_z\propto\tilde t_\perp^2\tau$, and Eq.\ (\ref{eq:sigma-tau})
reproduces AMRO.  However, for $\omega_c\tau\ll1$, electrons lose
coherence before they complete a cycle, so the interference effect is
washed out, and $\sigma_z$ reduces to $\sigma_z(0)\propto
t_\perp^2\tau$.  Fig.\ \ref{fig:bessel} shows $\sigma_z(\theta)$
calculated from Eq.\ (\ref{eq:sigma-tau}) for several values of
$\omega_c\tau$.  When $B$ is increased at a fixed angle $\theta$,
resistivity $\rho_z=1/\sigma_z$ increases and saturates at a finite
value in the limit $\omega_c\tau\to\infty$ for generic angles.
However, for the magic angles, $\rho_z$ increases without saturation,
because $\sigma_z\to0$ at $\omega_c\tau\to\infty$.  Notice that
observation of AMRO requires $\omega_c\tau>1$, whereas, according to
the Lifshitz-Kosevich formula, observation of the SdH oscillations
requires $\hbar\omega_c>T$, where $T$ is temperature.  These are
different conditions, and, typically, AMRO are still visible at
elevated temperatures, where the SdH oscillations have already
disappeared.  For example, in GaAs superlattices \cite{GaAs98}, AMRO
are clearly visible at 25 K, whereas the SdH oscillations dominate at
1.5 K.

Finally, we briefly discuss a possibility of observing AMRO in the
case where each layer of a bilayer is in the composite-fermion state
with the filling factor $\nu$ close to 1/2.  The composite fermions
experience the effective magnetic field $B_\perp^*=B_\perp(1-2\nu)$
and execute cyclotron motion with the radius $R_c^*=k_F^*\phi_0/2\pi
B_\perp^*$, where $k_F^*=\sqrt{2}k_F$ is their effective Fermi wave
vector.  By analogy, we would expect to see AMRO in the interlayer
conductivity with the magic angles given by Eq.\ (\ref{eq:theta_n})
with the substitution $B_\perp\to B_\perp^*$ and $k_F\to k_F^*$.
Unfortunately, the interlayer tunneling is greatly suppressed, because
the composite fermions need to decompose and recompose for tunneling
\cite{Spielman}.  However, the interlayer conductivity may increase at
higher temperatures and help to observe AMRO.  A systematic attempt to
observe AMRO would provide useful information about the nature of the
composite-fermion state.

\vspace{-1.5\baselineskip}


\begin{thebibliography}{99}
\itemsep 0pt


\bibitem{Kartsovnik88} M.V. Kartsovnik et al. JETP Lett. 48 (1988) 541.
  
\bibitem{Kajita} K. Kajita et al. Sol. St. Comm. 70 (1989) 1189.
  
\bibitem{Kartsovnik91} M.V. Kartsovnik et al. J. Phys. I (France) 2
  (1991) 89.

\bibitem{Kartsovnik96} M.~V. Kartsovnik, V.~N. Laukhin, J. Phys. I
  France 6 (1996) 1753.
  
\bibitem{Wosnitza96} J. Wosnitza et al. J. Phys. I France 6 (1996) 1597.
  
\bibitem{Singleton00} J. Singleton, Rep. Prog. Phys. 63 (2000) 1111.

\bibitem{Iye} Y. Iye et al. J. Phys. Soc. Jpn., 63 (1994) 1643.
  
\bibitem{Ohmichi} E. Ohmichi et al. Phys. Rev. B 59 (1999) 7263.
  
\bibitem{Dragulescu} A. Dragulescu et al. Phys.  Rev. B 60 (1999)
  6312.

\bibitem{Hussey} N.E. Hussey et al. Nature 425 (2003) 814.
  
\bibitem{GaAS91} R. Yagi et al. J. Phys. Soc.  Jpn. 60 (1991) 3784.
  
\bibitem{GaAs98} M. Kawamura et al. Physica B 249--251 (1998) 882.
  
\bibitem{GaAs01} T. Osada et al. Physica B 294--295 (2001) 402.

\bibitem{Yamaji} K. Yamaji, J. Phys. Soc. Jpn. 58 (1989) 1520.
  
\bibitem{Yagi} R. Yagi et al. J. Phys. Soc. Jpn.  59 (1990) 3069.
  
\bibitem{Kurihara92} Y. Kurihara, J. Phys. Soc. Jpn. 61 (1992) 975.
  
\bibitem{Kurihara93} Y. Kurihara, J. Phys. Soc. Jpn.  62 (1993) 255.
  
\bibitem{Yoshioka} D. Yoshioka, J. Phys. Soc. Jpn. 64 (1992) 3168.
  
\bibitem{McKenzie} R.H. McKenzie, P. Moses, Phys. Rev. Lett. 81 (1998)
  4492.
  
\bibitem{Moses} P. Moses, R.H. McKenzie, Phys. Rev. B 60 (1999) 7998.
  
\bibitem{Osada02} T. Osada, Physica E 12 (2002) 272.
  
\bibitem{Osada03} T. Osada et al. Physica E 18 (2003) 200.
  
\bibitem{Eisenstein} J.P. Eisenstein et al. Phys. Rev. B 44 (1991)
  6511.
  
\bibitem{Simmons93} J.A. Simmons et al. Phys. Rev. B 47 (1993) 15741.
  
\bibitem{Boebinger91} G.S. Boebinger et al. Phys. Rev. B 43 (1991)
  12673.
  
\bibitem{Harff} N.E. Harff et al. Phys. Rev. B, 55 (1997) R13405.
  
\bibitem{MacDonald92} J. Hu, A.H. MacDonald, Phys. Rev. B 46 (1992)
  12554.
  
\bibitem{Lyo93} S.K. Lyo, J.A. Simmons, J. Phys.: Condens. Matter 5
  (1993) L299.
  
\bibitem{Lyo98} S.K. Lyo, Phys. Rev. B 57 (1998) 9114.
  
\bibitem{Simmons98} S.K. Lyo et al. Phys. Rev. B 58 (1998) 1572.
  
\bibitem{Fertig02} R. C\^ot\'e et al. Phys. Rev. B 66 (2002) 205315.
  
\bibitem{Yakovenko} B.K. Cooper, V.M. Yakovenko, Phys. Rev. Lett. 96
  (2006) 037001.

\bibitem{Stark} C.B. Friedberg, R.W. Stark, Low Temp. Phys. (LT-13) 4
  (1974) 177.
  
\bibitem{ClO4} X. Yan et al. Synth. Metals 27 (1988) 145.
  
\bibitem{Spielman} I.B. Spielman et al. Phys. Rev. B 70 (2004) 081303.
  
\end{thebibliography}
\end{document}